\documentclass[doublecol,figures]{epl2}

\usepackage{epsf}
\usepackage{graphics}
\usepackage{subfigure}
\usepackage{amsmath}
\usepackage{amssymb}

\title{Localization behavior of vibrational modes in granular packings}

\author{Zorana Zeravcic\inst{1}\thanks{E-mail: \email{zorana@lorentz.leidenuniv.nl}} \and Wim van Saarloos\inst{1} \and David R. Nelson\inst{2}}
\shortauthor{Z.  Zeravcic \etal}

\institute{
  \inst{1} Instituut - Lorentz, Universiteit Leiden, Postbus 9506, 2300 RA Leiden, The Netherlands\\
  \inst{2} Department of Physics, Harvard University, Cambridge MA, 02138
}

\abstract{
We study the localization of vibrational modes of frictionless granular media. We introduce a new method, motivated by earlier work on non-Hermitian quantum problems, which works well both in the localized regime where the localization length $\xi $ is much less than the linear size $  L$ and in the regime $\xi \gtrsim L$ when modes are extended throughout our finite system. Our very lowest frequency modes show ``quasi-localized'' resonances away from the jamming point; the spatial extent of these regions increases as the jamming point is approached, as expected theoretically. Throughout the remaining frequency range, our data show no signature of the nearness of the jamming point and collapse well when properly rescaled with the system size. Using Random Matrix Theory we derive the scaling relation  $\xi \sim L^{d/2} $ for the regime $\xi \gg L $ in $d$ dimensions.
}

\begin{document}

\maketitle

\section{Introduction}

Over the past years, many questions concerning the behavior of disordered systems have been put in a new perspective  by addressing them from the point of view of the  more general jamming scenario \cite{liu1998}. Especially for granular systems it has turned out to be very fruitful to study the changes in the properties and the response of granular packings as one approaches the jamming point from the jammed side, where the packing gets close to an isostatic solid. An isostatic packing is indeed essentially a marginal solid which has just enough contacts to maintain a stable packing. From simple counting arguments, one finds that the average coordination number $Z$ of an $d$-dimensional isostatic packing of frictionless spheres equals $Z_{\rm iso}=2d$ \cite{alexander}. Upon approaching this marginal solid, many static and dynamic properties exhibit anomalous behavior, associated with the fact that the excess number of average bonds, $\Delta Z\equiv Z-Z_{\rm iso} $, goes to zero \cite{epitome,wyart,wyartlett,wyartE}. In fact,  $\Delta Z$ itself scales anomalously, namely as the square root of the difference in density from the one at jamming \cite{epitome}.  Likewise, the ratio G/K of the shear modulus G over the compression modulus K is found to scale as $\Delta Z$, and the density of states of the vibrational modes  becomes flat at low frequencies above some crossover frequency $\omega^* \sim \Delta Z$, due to the emergence of many low frequency modes. Much of this behavior was explained by  Wyart {\em et al.}\cite{wyart,wyartlett,wyartE} in terms of the existence of an important cross-over length scale $\ell^*\sim 1/\Delta Z$, the length up to which the response is close to that of an isostatic packing. This scale $\ell^*$ diverges as the jamming point is approached, but is difficult to probe directly. Nevertheless, the length $\ell^*$ has recently been uncovered as the important cross-over length to continuum behavior in the static response \cite{wouter,ellenbroek2008}. Although most of these results pertain explicitly to packings of frictionless spheres, there are several indications \cite{somfai2007,shundyak2007} that many of these observations and ideas can be generalized to frictional packings.

\begin{figure}
\onefigure[width=60mm]{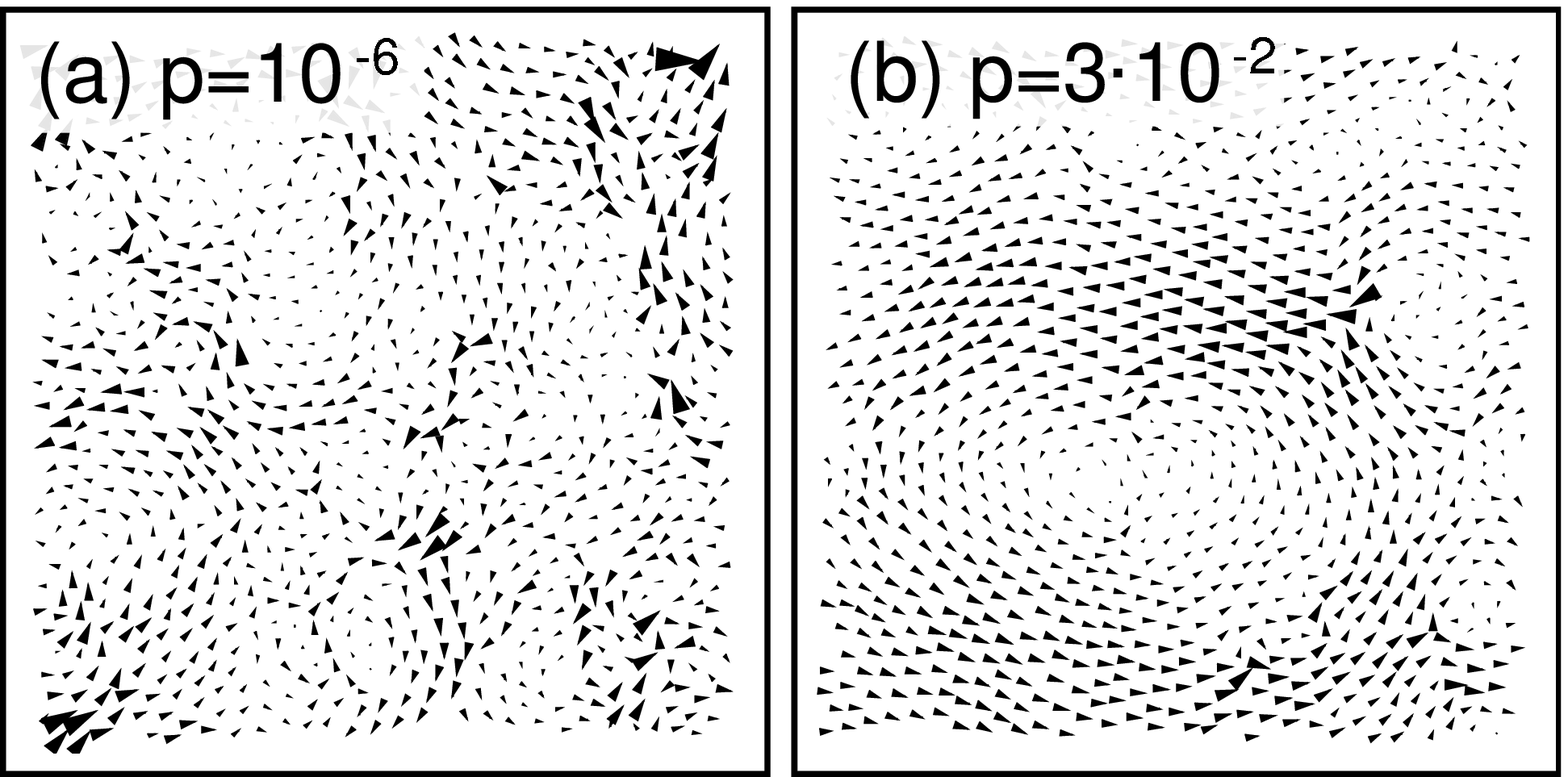}
\caption{Snapshots of two low-frequency eigenmodes in our packings. The arrows indicate the direction and magnitude of the displacements of the individual particles. (a) At low pressure $p=10^{-6}$, close to the jamming point, the mode is very disordered, whereas at high pressure (b) $p=3\cdot10^{-2}$, the mode is more reminiscent of an elastic shear wave. Similar features are seen in the response to a local or global deformation  \cite{wyart,wouter,tanguy}.}
\label{snapshots}
\end{figure}

It has been noted in several studies that both the response to a local or global deformation \cite{wouter,tanguy} and the behavior of the vibrational eigenmodes \cite{wyart,wyartE} of a packing become much more disordered as one approaches the jamming point: as the snapshots of two vibrational modes in Fig.~\ref{snapshots} illustrate, far above the jamming point the eigenmodes have a structure reminiscent of what one gets in a continuum theory of an elastic medium, but close to the jamming point one is immediately struck by the appearance of many disordered ``swirls''. The arguments put forward by Wyart {\em et al.} \cite{wyart,wyartlett,wyartE} indicate that the excess low frequency modes cannot be localized on scales $\lesssim \ell^*$ since they are the vestiges of the {\em global} floppy modes that emerge at the isostatic point. Hence, if there are any low-frequency modes away from jamming and if indeed their localization length is $\gtrsim \ell^*$, we should see this as the jamming point is approached. The aim of this paper is to investigate whether this is indeed the case.

Localization was discovered fifty years ago by Anderson\cite{anderson}, who  in his study 
of non-interacting electrons in a random potential found that disorder can induce electron localization. Unlike the extended (delocalized) Bloch waves, in a localized state the weight of the electron wave function is concentrated near some point in space; the amplitude falls off  as $e^{-r/\xi}$  with distance  $r$ from the center. This defines the localization length $\xi(E)$ which depends on the electron energy $E$. The possibility that disorder can localize the eigenmodes of systems governed by wave equations is quite general and extends to many systems, not only sound modes \cite{john,bunde,sheng} but also gravity waves \cite{sheng}, light propagation \cite{sheng} and diffusion on random lattices \cite{bunde,sheng}. We will focus on the localization behavior of vibrational modes of 2$d$ frictionless packings. In two dimensions there is no localization-delocalization transition: in the presence of disorder the states are generally localized in the thermodynamic limit  for cases like the one presented here \cite{john}.

The dynamic response of granular packings is affected by  three types of disorder --- bond disorder, mass disorder and topological packing disorder.  Any of these is sufficient to cause localization, but in practice all three play a role for realistic models of granular packings: bond disorder is present for all force laws except
one-sided harmonic springs, polydisperse particles will
have varying masses, and topological disorder is naturally present except for especially prepared regular piles, like a regular stack of marbles. Of course, in computer models these effects can be separated easily; we will not attempt to disentangle these three
contributions here, but do use this freedom
later
 to our advantage in testing our scaling predictions.

\begin{figure}
\onefigure[width=60mm]{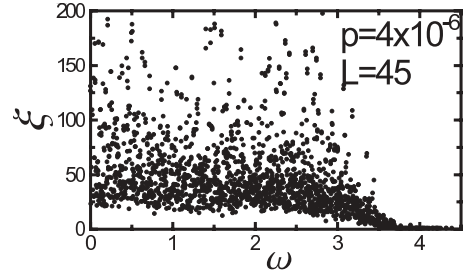}
\vspace{-0.2cm}
\caption{Scatter plot of the angularly averaged $\xi$'s of all the 2000 modes of our granular packing of 1000 particles  as a function of the frequency $\omega$ at pressure $p=4\cdot10^{-6}$ studied with the method explained in the text. Note the large scatter and the fact that the $\xi$ values are of order of the linear system size $L=45$ or larger throughout most of the frequency range.  }
\label{spread}
\end{figure}

\begin{figure}
\begin{center}
\includegraphics[width=25mm]{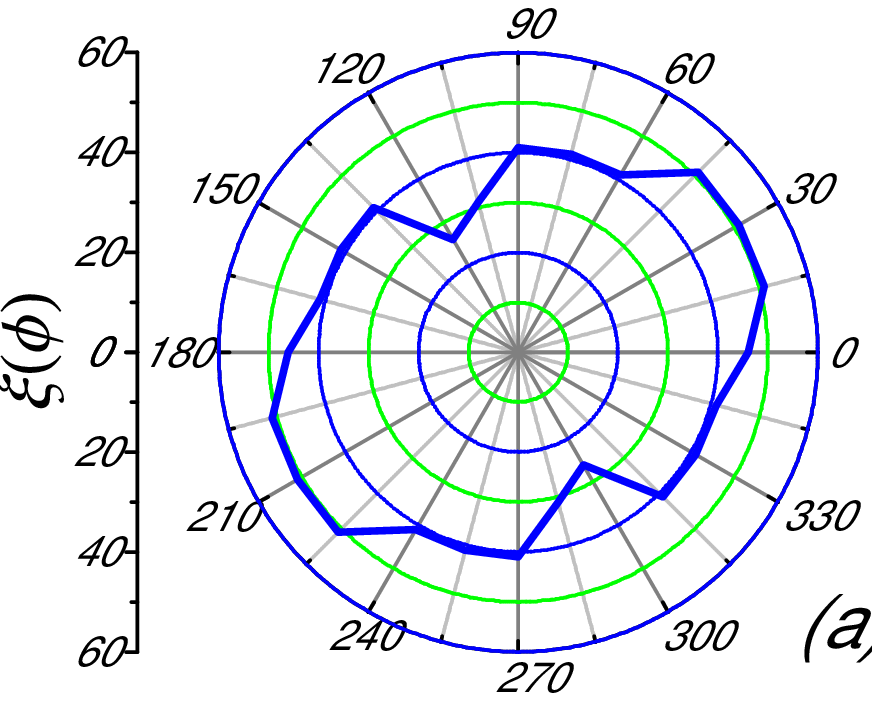}\hspace*{5mm}
\includegraphics[width=25mm]{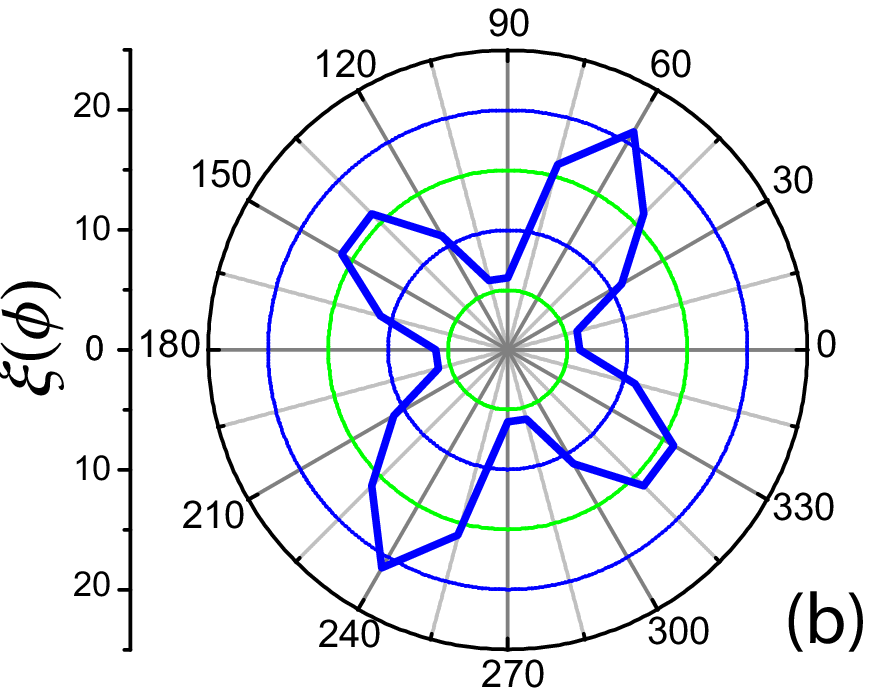}\hspace*{5mm}
\includegraphics[width=25mm]{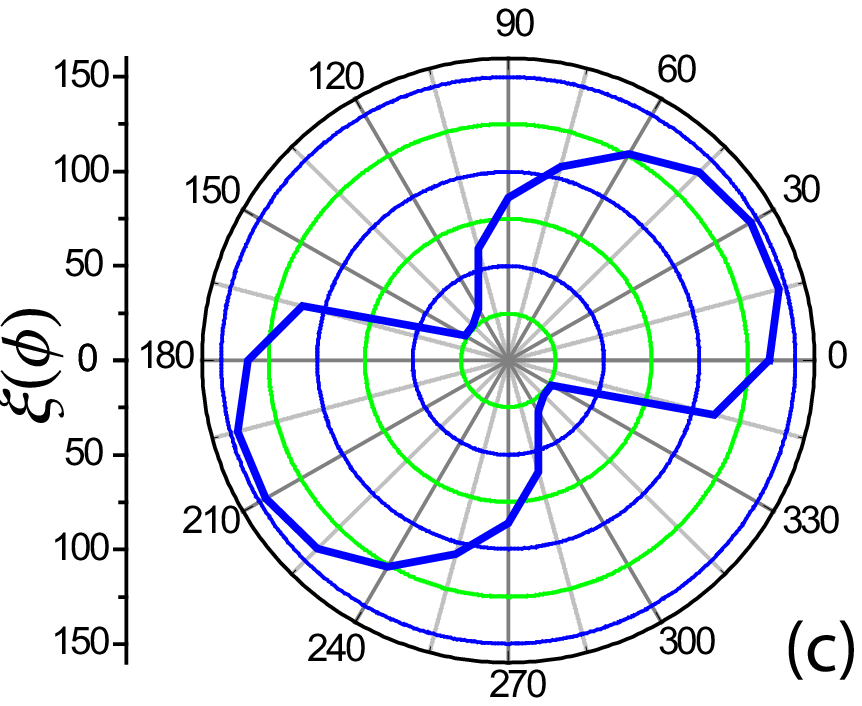}
\end{center}
\vspace{-0.5cm}
\caption{Polar plot of the localization length $\xi(\phi)$ in one of our granular packings at $p=4\cdot10^{-4}$ at low in (a), high in (b), and intermediate frequency in (c). The angular variation of $\xi(\phi)$ is comparable to the angularly averaged value itself. }
\label{angularplots}
\end{figure}

The crucial dilemma in extracting the localization length of the vibrational modes of granular packings  is that the effective disorder is so weak that one needs prohibitively large systems to reach  the true localization regime $\xi \ll L$ for most modes. Here $L$ is the linear system size. At the same time, existing methods which are based on spatial averages (like the direct expression based on the second moment of the eigenmode or the (Inverse) Participation Ratio method \cite{haake}) do not give much insight into the structure of the modes when $\xi$ approaches the system size $L$, \emph{i.e.}, for modes which are extended throughout the finite system. As Fig.~\ref{spread} illustrates, this is the relevant regime throughout most of the frequency range, as only the modes with the highest frequency $\omega$ are truly localized. The method we introduce in this paper, which is motivated by earlier work on non-Hermitian quantum problems \cite{nelson}, is based on studying the response to an asymmetric perturbation. It not only gives the proper localization length $\xi$ of each localized mode, but at the same time assigns a well-defined and precise direction-dependent value $\xi (\phi)$ to each mode, that spans through our finite system --- see Fig.~\ref{angularplots}. We stress that although we will follow common practice in referring to $\xi$ as the {\em localization} length even for $\xi \gtrsim L$, one should keep in mind that many modes {\em extend} throughout our finite periodic system, as both Figs.~\ref{spread} and \ref{angularplots} illustrate. As we shall show, this method does allow us to study the scaling with system size and disorder, and opens up the possibility to bring Random Matrix Theory \cite{haake,beenakker} to bear on this class of problems.

While all methods essentially yield the same localization length in the localization regime $\xi \ll L$, the extension of the concept of a localization length to the regime $\xi \stackrel{>}{\sim}  L$ depends on the method used and it is not a priori clear what $\xi$ in this regime pertains to. For our method,  one can however extract useful information about the large system limit from studying the regime $\xi \stackrel{>}{\sim}  L$. In conventional methods, one finds $\xi\approx L$ if the system size is too small. With our method we find a disorder-dependence too which can be used to extract quantitative estimates of the intrinsic localization length. As we will discuss in a forthcoming paper, this is simplest in one dimension where we predict and find a scaling $\xi \simeq A L^{1/2}$ in the regime $ \xi \stackrel{>}{\sim} L$. Since we expect a crossover to the localization regime when the \emph{intrinsic} localization length obeys $\xi_{int} \simeq L$, this to estimate the infinite size localization length from the small system data simply as $ \xi_{int} \simeq  A^2$. Preliminary analysis \cite{zz} indicates that this simple estimate works well in 1$d$, but we focus here on the behavior as function of the distance from the jamming point in 2$d$.

Our main results can be summarized as follows {\em (i)} The average localization length $\bar{\xi}(\omega)$ of granular packings is largely independent of the pressure, and hence of the distance from the jamming point. {\em (ii)} However, away from the jamming point there are a few ``quasi-localized'' low-frequency modes which disappear when approaching jamming. This behavior is qualitatively in accord with theoretical expectations for the change in behavior near the jamming point. {\em (iii)} In accord with what is expected on the basis of Random Matrix  Theory (RMT) \cite{haake,beenakker}, modes with $\xi\lesssim L$ are effectively noninteracting and the distribution of their level spacing is Poissonian, while modes with $\xi \gtrsim L$ show level repulsion: the level spacing follows the so-called Wigner surmise of RMT. {\em (iv) } In the regime $\xi \gtrsim L$, $\bar{\xi}(\omega)$ scales as $L^{d/2}$ and is inversely proportional to the disorder strength, in $d$ dimensions. {\em (v)} Due to level repulsion the distribution $P(\xi) $ falls off for large $\xi$ as $1/ \xi^3$.

\section{Method}
We use 2$d$ packings of 1000 frictionless particles which are prepared using molecular dynamics  simulations --- see \cite{wouter,somfai2007,shundyak2007} for the description of our algorithm that gently prepares packing at a target pressure and other details. The particles interact with the 3$d$ Hertzian force law, $f_{ij}\backsimeq\delta_{ij}^{3/2}$, where $\delta_{ij}$ is the overlap between particles $i$ and $j$.  The unit of length is the average particle diameter. Unless noted otherwise we here present results for our most extensive studies with  20\% polydispersity in the radii, but runs with different amount of polydispersity give similar results. The masses $m_i$ of the grains are taken proportional to $R_i^3$, corresponding to packing of spheres in 2$d$. The confining pressure, with which we tune the distance from the jamming point, is in the range $p\in(10^{-6},3\cdot 10^{-2})$ in the units of the Young modulus of the particles. We employ periodic boundary conditions in both directions. Our use of the 3$d$ Hertzian force law implies that the vibrational bonds $k_{ij}=df_{ij}/d\delta_{ij}\sim \delta_{ij}^{1/2} \sim p^{1/3}$ are disordered (they vary from bond to bond) and get weaker at smaller pressures. The natural frequency scale therefore goes down with pressure as $p^{1/6}$. As in  \cite{somfai2007}, when reporting our data we will therefore always rescale all frequencies $\omega$ with a factor $p^{-1/6}$, as to be able to compare data at different $p$.

\begin{figure}
\begin{center}
\includegraphics[width=30mm]{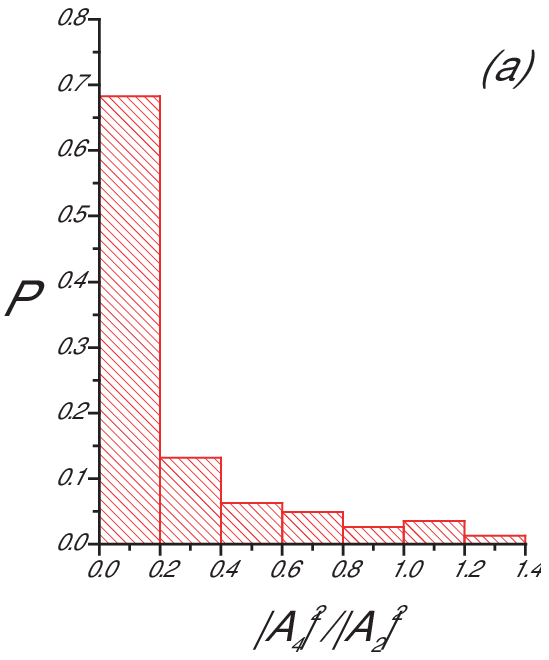}
\includegraphics[width=47mm]{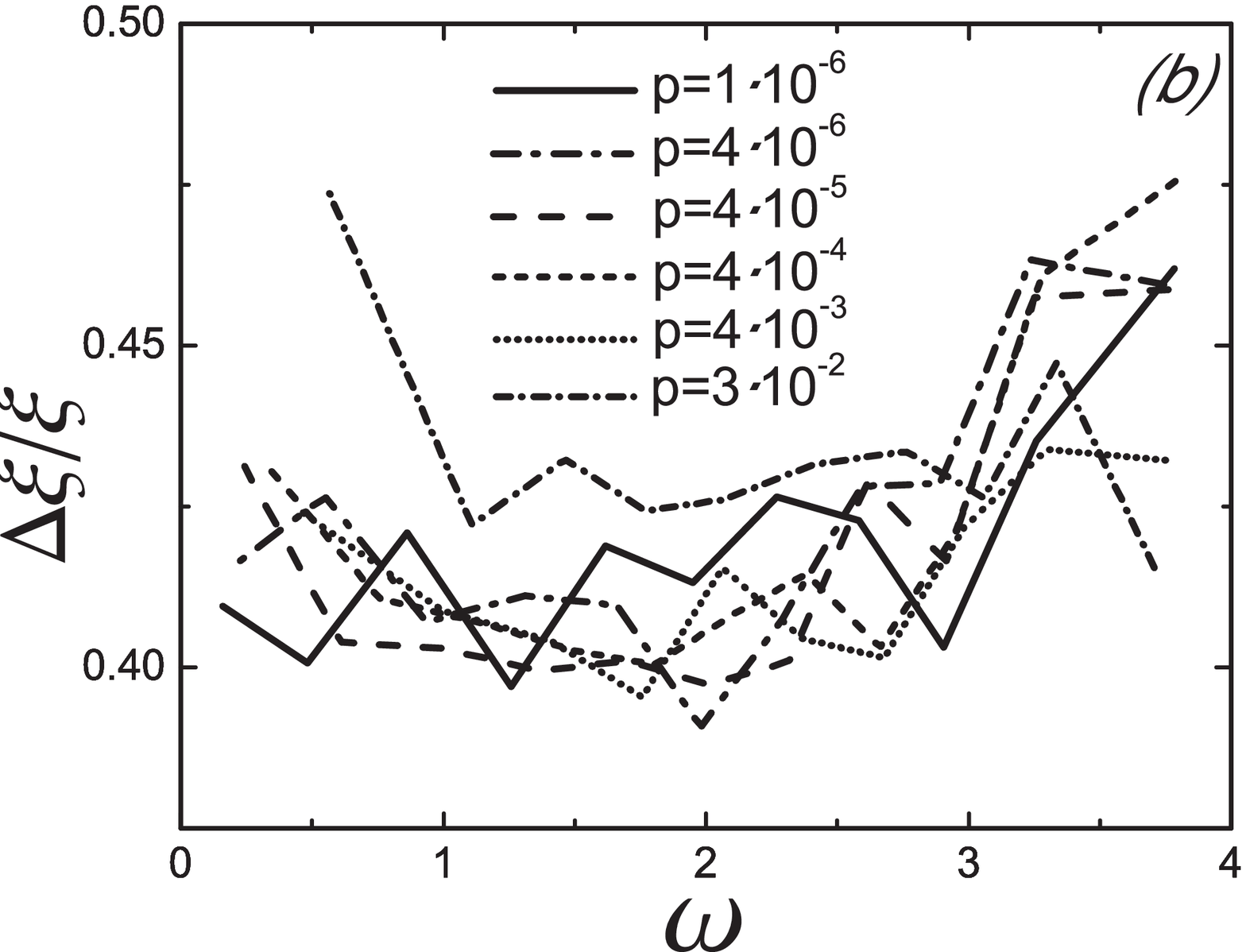}
\end{center}
\vspace{-0.2cm}
\caption{(a) Histogram of the ratio of squared amplitudes of the fourth (quadrupole) and second (dipole) harmonic at $p=4\cdot10^{-4}$. Most modes have predominantly dipole symmetry. (b) Average angular anisotropy $\Delta \xi / \bar{\xi} $ as a function of frequency for various pressures. }
\label{anisotropy}
\end{figure}

The vibrational modes and their density of states (DOS) are obtained in the standard way, by expanding the energy about the equilibrium positions of the grains up to quadratic terms. Just as in solid state physics, the dynamical  matrix, whose elements are the second derivatives of the energy with respect to the positions of the grains, determines the linear equations of motion of the vibrational modes. The dynamical matrix of a granular packing is a sparse symmetric matrix, because each particle only interacts with a few others.

Our method to extract the localization length is motivated by the work of Hatano and Nelson \cite{nelson} on the delocalization transition in non-Hermitian transfer matrix problems arising in the statistical mechanics of vortex lines in superconductors. Consider first the case of a one-dimensional chain of masses connected by springs with spring constants $k_{ij}$ ($j=i\pm1$) and periodic boundary conditions. We introduce an asymmetric bias term into the equations of motion so that the eigenvalue equation of a mode $u_i e^{-i\omega t}$ becomes
\begin{equation}
m_i \omega^2 u_i  =   \sum_{j=i\pm1}k_{ij} \left(e^{h\hat{x} \cdot \vec{x}_{ij}} u_{j}-u_i\right).
\end{equation}
Here $x_i$ are the rest positions of the particles and $\vec{x}_{ij}$ is a vector pointing from particle $i$ to particle $j$. For $h=0$ this is simply the dynamical equation for vibrations. The trick now is that we can extract the localization length $\xi_k$ of each mode $k$ by following whether or not its eigenvalue $\omega_k^2$ changes when we turn on $h$ in small steps. Indeed, as long as $h< 1/  \xi_k$ the eigenvalue $\omega^2_k$ will not change at all. To see this, note that in this case we can perform a  ``gauge transformation'' to a field  $\tilde{u}_i = {u}_i e^{h x_i}$  which obeys the original equation with $h=0$ {\em and} which falls off exponentially on both sides so that, in a large enough system, it  obeys the periodic boundary conditions. This implies that for  $h<1/\xi_k$, the eigenvalue $\omega^2_k$ does not change. However, once $h > \xi_k$ the function $\tilde{u}$ obtained with this transformation does not fall off exponentially to both sides. Thus, it can not obey  the periodic boundary condition with the same eigenvalue as it had for $h<1/\xi_k$: its eigenvalue {\em has} to change! In practice, when we increase $h$ the eigenvalue $\omega^2_k$ starts to change rapidly and collide with a neighboring eigenvalue when $h\approx 1/\xi$; beyond that, when $h\gtrsim 1/\xi_k$ the eigenvalue $\omega^2_k$ moves into the complex plane \cite{nelson}. Hence we can simply obtain the localization length $\xi_k$ of each mode $k$ from the value $h_k$ at which the eigenvalue moves into the complex plane upon increasing $h$: $\xi_k=1/h_k$. Note that in this method we do not need to calculate the eigenfunctions explicitly --- we only need to track the eigenvalues!

It is straightforward to extend this method to higher dimensions: as above, we simply multiply the off-diagonal elements of our dynamical matrix  with an exponential $e^{\vec{r}_{ij}\cdot\vec{h}}$, where $\vec{r}_{ij}$ is the vector pointing from the center of particle $i$ to its neighbor $j$. Our probe field $\vec{h}$ is now a vector, so by changing the angle that $\vec{h}$ makes with the $x$-axis, we can extract the angular anisotropy of the localization length $\xi(\phi)$ of each mode.

\section{Results}
\begin{figure}
\onefigure[width=68mm]{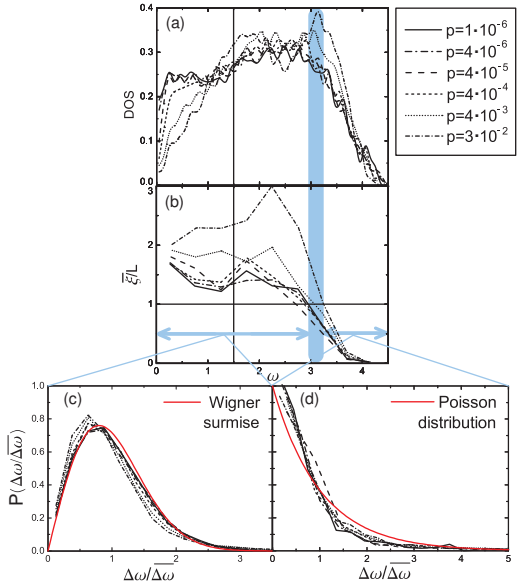}
\caption{(a) DOS of our packings for 6 different pressures confirming the main features of earlier studies  \cite{epitome,wyart,wyartlett,wyartE,somfai2007} close to jamming. (b) Our frequency binned and angularly averaged values $\bar{\xi}(\omega)/L$ are all very similar. (c,d)  Level spacing statistics  for the modes that have $\bar{\xi} \gtrsim L$ in (c) and for the modes with  $\bar{\xi}\lesssim L$. The lower frequency modes are essentially all extended and do show level repulsion in accord with the predictions from RMT \cite{haake,beenakker}, while the high frequency modes are truly localized and their level spacing is close to Poissonian. The gray lines indicate the frequency ranges used to obtain the level statistics in (c) and (d).}
\label{all}
\end{figure}

\begin{figure}
\onefigure[width=87mm]{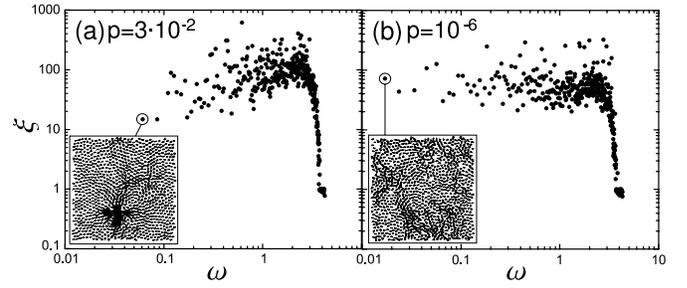}
\caption{Scatter plot for the localization lengths $\xi$ (determined to a precision of order unity) on a logarithmic frequency scale at $p=3 \cdot 10^{-2}$  in (a) and $p= 10^{-6}$ in (b) at system size $L=45$. Note that at the large pressure the lowest-frequency modes are localized; at small pressures this is not the case. The inset illustrates the two lowest-frequency modes, which has $\xi/L \approx 0.3$ in (a) and $\xi/L \approx 2$ in (b).}
\label{logarithmicplot}
\end{figure}

We first discuss some properties of the localization length of individual modes before turning to their scaling as a function of frequency, system size and distance from the jamming point.

{\em Anisotropy} --- Fig.~\ref{angularplots} shows the angular dependence $\xi(\phi) $ of three typical modes. One clearly sees that $\xi(\phi )$ is a $\pi$-periodic function and that the angular variation of $\xi(\phi)$ is significant. While few modes, like the second one in Fig.~\ref{angularplots}, have a quadrupolar structure, the anisotropy is predominantly dipolar, as the histogram in Fig.~\ref{anisotropy}(a) shows.

We will denote from here on the angularly averaged value of the localization length of an individual mode by $\xi$. Figure~\ref{anisotropy}(b) shows that  the root mean square average angular variation $\Delta \xi$  of $\xi(\phi)$ is almost half $\xi$, and that it is slightly larger at  higher  frequencies. There is no strong dependence of the anisotropy on the pressure, \emph{i.e.}, on the distance from the jamming point.

{\em Spread} --- The angularly averaged values $\xi(\omega)$ also show a large spread, as Fig.~\ref{spread} illustrates for a small value of the pressure. One also sees from this figure that most modes have a value of $\xi\gtrsim L$,  which means that they are extended within the systems we can analyze --- only our largest frequency modes are truly localized \cite{privatecommunication,theirresults}.

{\em DOS} --- We now turn to a more systematic analysis of our data as a function of pressure and system size. In Fig.~\ref{all}(a) we show that the  density of states (DOS) of our packings behaves as found before \cite{epitome,wyart,wyartlett,wyartE,somfai2007} for such packings: As the the jamming point is approached by lowering the pressure, the density of low-frequency modes increases dramatically, which, as mentioned before, is due to the nearness of the isostatic point.

{\em Average localization length $\bar{\xi}(\omega)$ } ---
For each dataset of the individual angularly averaged values of $\xi$, as in Fig.~\ref{spread}, we determine the frequency binned average values $\bar{\xi}(\omega)$ (each based on about 100 to 200 modes). The behavior of $\bar{\xi}/L$ as a function of (scaled) frequency is show in Fig.~\ref{all}(b) for six different values of the pressure. In these average values, there is no strong variation with pressure, \emph{i.e.} with distance to jamming.

We already noted in Fig.~\ref{spread} that most of our eigenmodes have $\xi \gtrsim L$, \emph{i.e.} are extended in our finite system. This is also clear from Fig.~\ref{all}(b): at all but the largest frequencies we have, $\bar{\xi}\gtrsim L$. There are indeed roughly three regimes present in Fig.~\ref{all}(b). From high frequencies towards low frequencies, we first have a range of high-frequency localized modes, for which $\bar{\xi}<L$. These modes are always present at any pressure and are the high-frequency modes in which only a few (light) particles oscillate more or less in anti-phase as in an optical mode (such type of modes generally arise immediately when disorder is introduced into an ordered system). For intermediate-range frequencies there is  a plateau in $\bar{\xi}$.  Finally for the lowest frequencies (in the frequency range where actually the excess modes appear in  the  DOS in Fig.~\ref{all}(a) at low pressures), there is an indication of an upswing in $\bar{\xi}$ for small $\omega$. We find this upswing at low frequencies in all our data, also on percolation lattices \cite{zz}, where it is even more pronounced.

{\em Quasi-localized low-frequency modes at high pressure} --- From the above data for the bin-averaged $\bar{\xi}$, it would appear at first sight that we see no signature of the nearness of the jamming point. This, however, is not true: in Fig.~\ref{all} we show data obtained by averaging over 100-200 modes. However, this averaging washes out systematic trends visible for the lowest frequency eigenmodes discovered by Vitelli, Xu \emph{et al.} \cite{theirresults,privatecommunication}. When plotted on a logarithmic scale, as in Fig.~\ref{logarithmicplot}, we see a systematic trend for $\xi$ of the low frequency modes to decrease with increasing pressure. As the inset of Fig.~\ref{logarithmicplot}(a) illustrates, these are ``quasi-localized'' modes in which a reasonably well defined ``localized'' group of particles performs what looks like a resonant oscillation that is weakly coupled to the extended elastic field. For our limited range of $L$, we find $\xi/L\simeq0.3$ and a reduced anisotropy of $\Delta \xi/\xi \approx 0.2$ for these modes.

\begin{figure}
\begin{center}
\includegraphics[width=41mm]{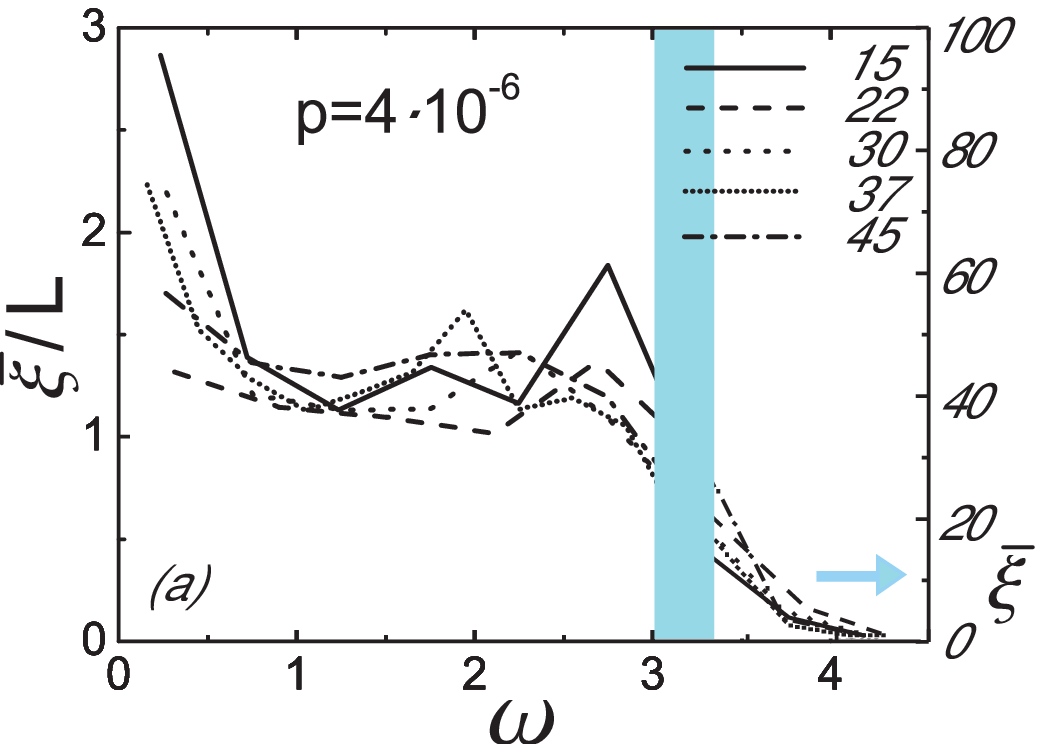} 
\includegraphics[width=43mm]{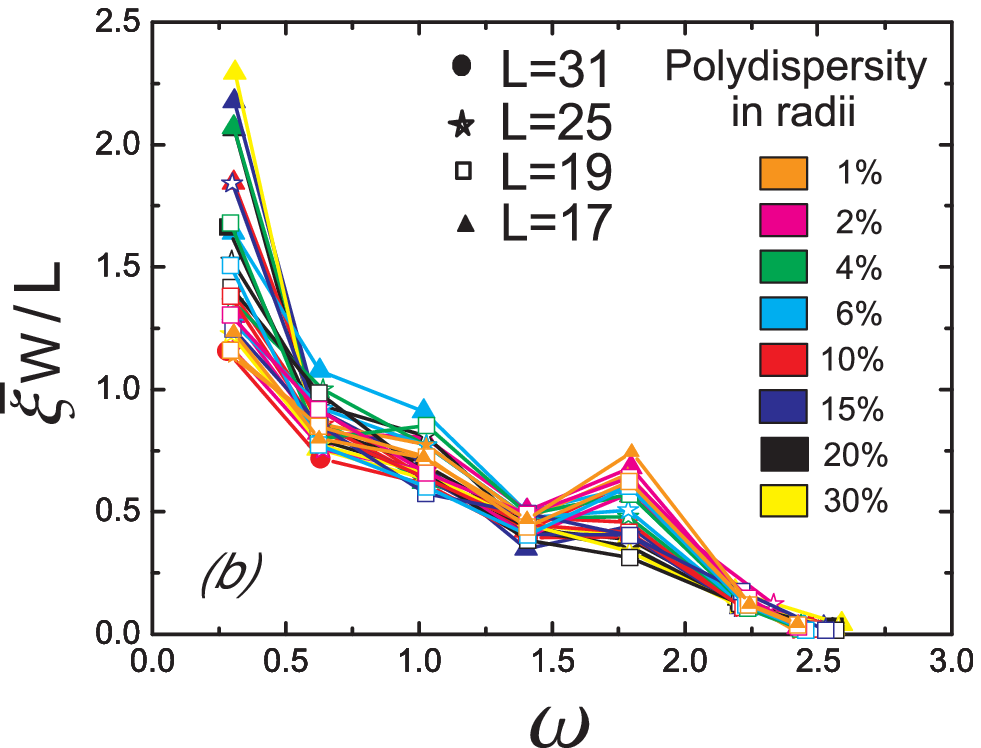}
\end{center}
\vspace{-0.3cm}
\caption{(a) Finite size scaling for $p=4\cdot10^{-6}$ and linear system size $L$ ranging from 15 to 45, confirming that the extended states, where $\bar{\xi}\gtrsim L$, scale with $L$, while the high-frequency modes are, within the statistical error, $L$-independent. (b) Scaling collapse  of $\bar{\xi}$ according to (\ref{ourscaling}) for hexagonal lattices with identical springs but varying masses $m_i \propto R_i^3$, as for spheres. The distribution of radii $R_i$ is taken to be flat, and  $W$ is taken to be the width of the distribution in percent. }
\label{fss}\label{hexagonalresults}
\end{figure}

As we discussed in the introduction, for  packings closer to the jamming point (at  lower pressures) the isostaticity length $\ell^*$ increases as $1/\Delta Z$, where $\Delta Z$ is the excess contact number. Up to this scale $\ell^*$ we do not expect localized modes at low frequencies, since up to this scale the response  mirrors that of the {\em global} floppy modes that emerge near the isostatic point. Indeed, within the system sizes we can study there are no low-frequency ``quasi-localized'' modes at all at low pressures, as Fig.~\ref{logarithmicplot}(b) illustrates for $p=10^{-6}$, even though the response is in many ways more disordered due to the nearness of the jamming point!

While our data are qualitatively in accord with the above scenario, we have unfortunately too few low-frequency ``quasi-localized'' modes to confirm quantitatively that as we tune the packings closer to jamming the extent of the resonant region increases with $ \ell^* \sim 1/ \Delta Z$.

{\em Level Spacing Statistics } --- Based on the results of RMT \cite{haake,beenakker}, one expects the following: the frequencies $\omega_i$ of the localized modes should be independently distributed, so that their spacing $\Delta \omega_i=\omega_{i+1}-\omega_i$ obeys a Poisson distribution, while the modes which extend throughout the system should interact and repel each other, with a level spacing distribution which is given by the Wigner surmise, $P_W(s)=\pi s/2\exp{(-\pi s^2/4)}$, where $s=\Delta \omega / \overline{\Delta \omega}$. Figs.~\ref{all}(c,d) confirm that this expectation is fully born out by our data at all pressures. Note that the distribution in Fig~\ref{all}(c) deviates somewhat from the Wigner surmise at the two highest pressures --- this is due to the ``quasi-localized'' low-frequency modes discussed above.

{\em Scaling with system size $L$} --- One would of course expect $\bar{\xi}$ for the modes which extend through our system size to scale with $L$. As we will sketch below, we have used RMT \cite{haake,beenakker} to derive for our method the scaling behavior $\xi \sim L^{d/2}$. More generally we propose
\begin{equation}
\bar{\xi} \sim L^{d/2}/W,  \label{ourscaling}
\end{equation}
where $W$ is a measure of the effective disorder.
Fig.~\ref{fss}(a) shows that the $\bar{\xi}\sim L$-scaling is well obeyed for our two-dimensional packings for the extended modes in the range $\omega \lesssim 3$ (as noted before the quasi-localized modes obey $\bar{\xi}\simeq 0.3L$), while the high-frequency localized modes for $\omega \gtrsim 3.4$ have $\bar{\xi}$'s which are indeed essentially $L$-independent.

For our gently prepared granular packings the strength of the disorder can not easily be varied. In order to test our scaling prediction (\ref{ourscaling}), we have prepared {\em ordered} hexagonal lattices with all spring constants the same but varying masses. As Fig.~\ref{hexagonalresults}(b) shows, we obtain very good data collapse with (\ref{ourscaling}) at all but the highest frequencies. Note also that for small amount of disorder, we have $\bar{\xi} \gg L$. Results for one-dimensional chains are fully consistent with the predicted $L^{1/2}$ scaling \cite{zz}.

\section{Implications from Random Matrix Theory} Let us finally summarize what Random Matrix Theory (RMT) brings to bear on the study of the localization length. We refer for a more extensive discussion to \cite{zz}.

{\em 1)} In RMT it is well known that for analyzing the level statistics, like  in Fig.~\ref{all}(c,d), it is important to use the proper variable. The procedure to obtain the proper variable is the so-called ``unfolding of the spectrum'' \cite{haake}. For our case, the unfolding ensures that in each small interval, the mean level spacing is the same as in the original spectrum. The proper variables are then indeed the frequencies $\omega$, not the eigenvalues $\omega^2$ of the dynamical matrix.

{\em 2)} The scaling (\ref{ourscaling}) of the modes with $\xi \gtrsim L$ can be understood as follows. When we turn on $h$, energy levels start to move on the real axis, some getting closer together and some further apart. Because of reflection symmetry under $\vec{h} \to - \vec{h}$ (which is also apparent in Fig.~\ref{angularplots}), the shift is quadratic in $h$. We determine $\xi$ of a mode from the collision value $h_{\rm c} $ at which two modes collide along the real $\omega$-axis  and ``pop'' into the complex plane.  According to RMT \cite{haake}, the typical collision parameter is then  $h_{c}^2\approx \frac{mean\ level\ spacing}{typical\ level\ velocity}$. For our systems the mean level spacing is proportional to $ 1/L^{d}$ and the typical level velocity does not depend on $L^{d}$, from which the scaling $\bar{\xi} \sim L^{d/2}$ immediately results upon identifying $h_c$ with $\xi^{-1}$.

{\em 3)} In line with the large spread in the values of $\xi$, we find that the distribution  $P(\xi /\bar{\xi})$ implied by Fig.~\ref{spread} falls off as $(\xi /\bar{\xi})^{-3}$ for large $\xi$ both in 1 and 2 dimensions. This power law decay can be derived from how the level repulsion of the extended modes changes, when we change the perturbation parameter $h$ by $\Delta h$ \cite{zz}.

\section{Conclusions and Outlook}
In this paper we have introduced a new method, motivated by previous studies of non-Hermitian quantum problems \cite{nelson}, which allows an analysis of localization in phonon spectrum, including the regime $\bar{\xi}\gtrsim L$ when the eigenmodes are extended within the finite systems we can study. The method is especially relevant for granular packings, where $\bar{\xi}\gtrsim L$ throughout most of the frequency range, since even in this regime our method gives different results depending on the amount of disorder.  For the system sizes that are numerically accessible at present we can not yet test the proposed scaling relation $\xi \gtrsim \ell^* \sim 1/ \Delta Z$ quantitatively. Nevertheless, the disappearance of the ``quasi-localized'' low-frequency modes as we approach the jamming point by lowering the pressure, agrees with the scenario advanced by Wyart {\em et al.} \cite{wyart,wyartlett,wyartE} that up to this length scale the low-frequency rearrangements and modes extend over a diverging scale $\ell^*$. We aim to study larger systems and more packings in the future using sparse matrix eigenvalue routines rather than a Mathematica program, to investigate the nature and scaling of the low-frequency modes in more detail.

A few final remarks are in order. {\em (i)} Our method will allow us to determine which type of disorder (mass disorder, bond disorder or topological disorder) plays the dominant effect in the localization behavior. {\em (ii)} The resonant region of the quasi-localized mode shown in Fig.~\ref{logarithmicplot} has quadrupolar symmetry, and is reminiscent of the quadrupolar deformation fields that have been proposed \cite{lemaitre} to dominate quasistatic shear relaxation. Although this is not true for all quasi-localized modes, this may not be accidental. The possible connection between these resonances and shear transformation zones  is extremely intriguing and should be pursued further. {\em (iii)}  The results for $\bar{\xi}(\omega)$ in  finite system typically show an upswing  for small $\omega $, except at the largest pressures; whether this is a finite system analogue of the  well known $\omega \to 0$ divergence in infinite 2$d$ systems \cite{john} is unclear to us. {\em (iv)} The states with large but finite localization lengths at low frequency that we find at high pressures (see Fig.~\ref{logarithmicplot}(a)) are intriguing. It will be interesting to see if these states persist in the presence of the entropic interactions at finite temperature. {\em (v)} Diffusion on percolation lattices is also an appealing model system to apply the method to:  close to the percolation threshold most eigenmodes are  truly localized and thus have $\xi \ll L$, while away from the percolation threshold there is a crossover to the regime where $\bar{\xi} \gg L$  \cite{zz}.

\acknowledgments
We are grateful to Sid Nagel, Andrea Liu and Vincenzo Vitelli for illuminating discussions and for stressing to focus on the the lowest frequency modes, to Jens Bardarson, Martin van Hecke, Kostya Shundyak and Dani ben-Avraham for their interest and advice, and Wouter Ellenbroek and Ell\'{a}k  Somfai for supplying the granular packings needed for this study. DRN would like to acknowledge conversations with Bertrand I.~Halperin. ZZ acknowledges support from physics foundation FOM, and DRN support of the National Science Foundation, through Grant DMR-0654191 and the Harvard Materials Research Science and Engineering Center, through Grant DMR-0213805.

\end{document}